\documentclass[aps,prl,twocolumn,notitlepage,superscriptaddress]{revtex4-1}

\usepackage[utf8]{inputenc}
\usepackage{amsmath,amsfonts,amssymb}
\usepackage{graphicx}

\usepackage{hyperref}
\usepackage[usenames,dvipsnames]{xcolor}

\usepackage{tikz}
\hypersetup{colorlinks=true, linkcolor=BrickRed, urlcolor=blue!50!black, citecolor=blue!50!black, pdfproducer={}}

\renewcommand{\vec}[1]{\mathbf{#1}}
\renewcommand{\phi}[0]{\varphi}

\usepackage[normalem]{ulem}

\begin{document}

\title{Target search of active agents crossing high energy barriers}

\author{Luigi Zanovello}
\affiliation{Institut f\"ur Theoretische Physik, Universit\"at Innsbruck, Technikerstra{\ss}e 21A, A-6020, Innsbruck, Austria}
\affiliation{Dipartimento di Fisica, Universit{\`a} degli studi di Trento, Via Sommarive 14, 38123 Trento, Italy}
\author{Michele Caraglio}
\affiliation{Institut f\"ur Theoretische Physik, Universit\"at Innsbruck, Technikerstra{\ss}e 21A, A-6020, Innsbruck, Austria}
\author{Thomas Franosch}
\email{Thomas.Franosch@uibk.ac.at}
\affiliation{Institut f\"ur Theoretische Physik, Universit\"at Innsbruck, Technikerstra{\ss}e 21A, A-6020, Innsbruck, Austria}
\author{Pietro Faccioli}
\email{pietro.faccioli@unitn.it}
\affiliation{Dipartimento di Fisica, Universit{\`a} degli studi di Trento, Via Sommarive 14, 38123 Trento, Italy}
\affiliation{INFN-TIFPA, Via Sommarive 14, 38123 Trento, Italy}

\date{\today}

\begin{abstract}
Target search by active agents in rugged energy landscapes has remained a challenge because standard enhanced sampling methods do not apply to irreversible dynamics.
We overcome this non-equilibrium rare-event problem by developing an algorithm generalizing transition-path sampling to active Brownian dynamics.
This method is exemplified and benchmarked for a paradigmatic two-dimensional potential with a high barrier.
We find that even in such a simple landscape the structure and kinetics of the ensemble of transition paths changes drastically in the presence of activity. Indeed,  active Brownian particles reach the target
more frequently than passive Brownian particles, following longer and counterintuitive search patterns.
\end{abstract}

\maketitle

Active matter and directed motion are receiving increasing attention because of their relevance in a wide range of research fields, including biology, biomedicine, robotics, and statistical physics~\cite{bech2016,marc2013}.
Active propulsion allows bacteria and animals to explore their local environment and forage nutrients~\cite{bech2016,elge2015} and is key to the development of new nanoparticles that may act as drug delivery agents~\cite{naah2013,patr2013,chea2014,liu2016}.
Furthermore, phagocytes of the immune system perform chemotactic motion during injury or infection~\cite{devr1988,deol2016} and sperm cells navigate against chemical gradients to find the egg~\cite{eise2006}.
The central question in all these examples is how active agents find their target.
In spite of its relevance, to date this problem has been addressed by relatively few theoretical studies.
In particular, only the special case of run-and-tumble motion~\cite{volp2017} was investigated within the framework of intermittent search patterns~\cite{beni2011,visw1999,beni2005,beni2006,visw2008}.

Target search crucially depends on the environment~\cite{volp2017}, and in many realistic scenarios it involves exploring a complex energy landscape, characterized by the presence of several local minima, separated by energy barriers.
Unfortunately, the computational cost of simulating the dynamics by directly integrating the equations of motions grows exponentially with the ruggedness of the landscape and the height of the barriers.  

For passive systems, similar problems have been solved by the development of enhanced sampling methods~\cite{torr1977,Laio2002,bowm2013,dell1998,bolh2002,dell2003,verp2003,fara2004,bell2015,alle2009,huss2020} but much less attention has been devoted to enhanced sampling applications on active particles, with the notable exceptions of a case study of alignment interactions in a modified Vicsek model~\cite{buij2020} and of systems displaying motility-induced phase separation~\cite{redn2016,whit2018}.
Among the various enhanced sampling algorithms, transition-path sampling (TPS)~\cite{dell1998,bolh2002,dell2003} has the advantage to provide a completely rigorous sampling of the reactive trajectories between a reactant basin and a target basin.
This algorithm is essentially a Metropolis Monte Carlo performed in the space of reactive trajectories.
Trial moves are generated by choosing a random state on a known reactive path and integrating the equations of motion (shooting) forward and backward in time to obtain a new transition path.	
Trial moves are then accepted or rejected according to some probability explicitly calculated from the equations of motion.
The main problem encountered when applying TPS to active target search concerns the possibility to apply backward shooting.
In passive systems, this is possible, because the backward probability is directly related to forward dynamics by microscopic reversibility.
The dynamics of a self-propelled particle, however, is microscopically irreversible, therefore it is commonly believed that backward shooting is not allowed.
To overcome this problem, other methods such as forward flux sampling (FFS)~\cite{alle2009,huss2020} have been specifically developed to account also for non-equilibrium systems.
Nevertheless, these methods suffer other limitations, such as an efficiency drop due to the impossibility to shoot backward in time and the need to know {\it a priori} a proper reaction coordinate, which could be not always possible on complex energy landscapes.
Recently, two novel methods~\cite{buij2020} addressed this challenge by bridging between the original TPS and the FFS methodology.
As TPS, they aim at importance sampling of trajectory space and are independent of the choice of a reaction coordinate.
Yet, as FFS, they allow only forward shooting and may therefore be inefficient in certain situations.

Here, we show for the first time that in the case of an active Brownian particle (ABP) the lack of microscopic reversibility can be circumvented and backward shooting may therefore become feasible.
The result is a generalized version of the original TPS algorithm, in which the acceptance probability contains a new term depending explicitly on the particle's activity.
We then apply our new scheme to study how the activity affects target search in a two-dimensional landscape, characterized by the presence of a large energy barrier.

We first consider an ensemble of microswimmers initially confined to some reactant region R, searching for some target region T.
A single microswimmer is modeled as an ABP in two dimensions, i.e. the equations of motion consist of a set of Langevin equations in which the activity is provided by a stochastic drive proportional to a velocity term of modulus $v$ and subject to a rotational diffusion process.
If the ABP diffuses in a conservative energy landscape $U(x,y)$, the equations of motion discretized according to the It\^{o} rule are
\begin{eqnarray}\label{eom}
\vec{r}_{i\!+\!1} &=& \vec{r}_{i} + v\, \vec{u}_{i} \, \Delta t - \mu \vec{\nabla} U(\vec{r}_{i}) \Delta t + \sqrt{2D\Delta t} \, \boldsymbol{\xi}_i,\\ \label{eom2}
\vartheta_{i\!+\!1} &=& \vartheta_{i} + \sqrt{2D_{\vartheta}\Delta t} \, \eta_i,
\end{eqnarray}
where  $\Delta t$ is the integration step, $\vec{r}_i = (x_i,y_i)$ is the position at time $i \Delta t$ and $\vec{u}_{i} = \big(\cos\vartheta_{i},\sin\vartheta_{i}\big)$ denotes the instantaneous orientation of the driving velocity. 
$D$ and $D_{\vartheta}$ are the translational and rotational diffusion coefficients, respectively, and $\mu$ is an effective mobility~\cite{specification_passive_case}.
Finally, the components of the vector noise $\boldsymbol{\xi}_i=(\xi_{x,i},\xi_{y,i})$ and of the scalar noise $\eta_i$ are independent random variables, distributed according to a Gaussian with zero average and unit variance.
In the following we refer to the microstate of a single active particle as  $w=(\vartheta, \vec{r})$.

To sample the reactive paths from R to T we reconsider the TPS algorithm~\cite{bolh2002,dell2003} and adapt it to the present case of ABPs.
In TPS, a Markov chain of reactive trajectories is generated starting from some arbitrary initial path.
Trial moves (newly attempted transition paths $\mathcal{W}^{\text{new}}$ generated starting from an old path $\mathcal{W}^{\text{old}}$) are proposed according to a three-step procedure: 
First, a microstate $w_i^{\text{old}}$ is randomly picked from the frames in $\mathcal{W}^{\text{old}}$.
Next, the microstate may be modified by means of some random perturbation: $w_i^{\text{old}} \!\rightarrow\! w_j^{\text{new}}$.
Finally, a new trial trajectory $\mathcal{W}^{\text{new}}$ is obtained by shooting forward and backward in time, starting from $w_j^{\text{new}}$ (see Supplemental Material~\cite{supplement_TPS_ABP} -- SM). 
The resulting new trajectory $\mathcal{W}^{\text{new}}$ is then accepted with a  probability $\mathcal{P}_{\text{acc}}$, which is calculated from the underlying microscopic dynamics by imposing the detailed balance condition in the functional space of reactive trajectories.
This results in a standard Metropolis rule
\begin{equation} \label{eq_accprob}
\begin{split}
\mathcal{P}_{\text{acc}} & \left[\mathcal{W}^{\text{old}} \!\rightarrow\! \mathcal{W}^{\text{new}}\right] = h\left[\mathcal{W}^{\text{new}}\right] \times \\ &  \min \Bigg\{1,\frac{\mathcal{P}\left[\mathcal{W}^{\text{new}}\right]\mathcal{P}_{\text{gen}}\left[\mathcal{W}^{\text{new}}\!\rightarrow\! \mathcal{W}^{\text{old}}\right]}{\mathcal{P}\left[\mathcal{W}^{\text{old}}\right]\mathcal{P}_{\text{gen}}\left[\mathcal{W}^{\text{old}}\!\rightarrow\! \mathcal{W}^{\text{new}}\right]}\Bigg\} \; ,
\end{split}
\end{equation}
where $h\left[\mathcal{W}^{\text{new}}\right]$ is a characteristic function equal to one only if the new path is reactive and zero otherwise.
$\mathcal{P}[\mathcal{W}]$ is the functional path probability density which reads
\begin{equation} \label{eq_forpath}
\mathcal{P}\left[\mathcal{W}\right] \propto \rho(w_{0}) \prod_{i=0}^{N-1}p(w_{i}\! \rightarrow\! w_{i\!+\!1}) \, ,
\end{equation}
where $N$ is the number of frames of the reactive path $\mathcal{W}$, and  $\rho(w_0)$ is the  quasi-stationary non-equilibrium distribution of initial conditions in the reactant.
Further,
\begin{equation} \label{eq_forwardprob}
\begin{split}
 p(w_{i}\! & \rightarrow\! w_{i\!+\!1})  \propto \exp\Big\{\!-\frac{(\vartheta_{i\!+\!1}-\vartheta_{i})^2}{4D_{\vartheta}\Delta t}\Big\} \times \\ & \exp\Big\{\!-\frac{\left(\vec{r}_{i\!+\!1}-\vec{r}_{i}-v\vec{u}_{i} \Delta t +\mu \vec{\nabla}U(\vec{r}_{i}) \Delta t \right)^2}{4D\Delta t}\Big\},
\end{split}
\end{equation}
is the conditional probability for performing a transition from $w_i$ to $w_{i+1}$ in the infinitesimal time interval $\Delta t$ as is readily derived from Eq.~\eqref{eom} (see  SM).
$\mathcal{P}_{\text{gen}}\left[\mathcal{W}\right]$ in Eq.~\eqref{eq_accprob} is the probability of generating a trial path, according to the shooting procedure outlined above.
Explicitly,
\begin{equation} \label{eq_genprob}
\begin{split}
\mathcal{P}_{\text{gen}} & \left[\mathcal{W}^{\alpha} \! \rightarrow\! \mathcal{W}^{\beta}\right]  =  \mathcal{P}_{\text{sel}}(w_{j}^{\alpha}\! \mid\! \mathcal{W}^{\alpha} ) \mathcal{P}_{\text{pert}}(w_{j}^{\alpha}\! \rightarrow\! w_{i}^{\beta}) \times \\ & \qquad \prod_{k=i}^{N^{\beta}-1}p(w_{k}^{\beta}\! \rightarrow\! w_{k\!+\!1}^{\beta}) \prod_{k=0}^{i-1}\bar{p}(w_{k\!+\!1}^{\beta}\! \rightarrow\! w_{k}^{\beta}) \; ,
\end{split}
\end{equation}
where $\mathcal{P}_{\text{sel}}(w_{j}^{\alpha}\! \mid\! \mathcal{W}^{\alpha}) = 1/N^{\alpha}$ is the probability of selecting as shooting point the state $w_{j}^{\alpha}$ belonging to the reactive path $\mathcal{W}^{\alpha}$ of length $N^{\alpha}$.
$\mathcal{P}_{\text{pert}}(w_{j}^{\alpha}\!\rightarrow\!w_{i}^{\beta})$ is the probability of perturbing $w_{j}^{\alpha}$ to obtain the microstate $w_{i}^{\beta}$ of the path $\mathcal{W}^{\beta}$ of length $N^{\beta}$~\cite{length_paths,shooting_point}.
The first and second product of probabilities in the second line of Eq.~\eqref{eq_genprob} represent the probability of generating the two branches of the trial trajectory $\mathcal{W}^{\beta}$ connecting the microstate $w_{i}^{\beta}$ to the target and reactant state, respectively.
In particular, $\bar{p}(w_{i\!+\!1} \! \rightarrow\! w_i)$ is the probability to observe a transition from the microstate $w_{i\!+\!1}$ to the microstate $w_i$ in a dynamics evolving backwards in time.
For passive Brownian dynamics, this backward probability is directly related to the forward one by microscopic reversibility.
However, the dynamics of ABPs is intrinsically irreversible~\cite{dabe2019} and the calculation of $\bar{p}(w_{i\!+\!1} \! \rightarrow\! w_i)$ is non-trivial.
In principle, one could use any sort of backward dynamics, and consistently derive the corresponding acceptance probability within the TPS approach by following the procedure explained below.
However, the challenge is to find moves that are efficient in sampling the reactive paths, which is possible if the trial moves have a significant overlap with the ``real dynamics''. 
Here we suggest for the backwards shooting the simple rule:
\begin{eqnarray}\label{beom}
\vec{r}_{i} &=& \vec{r}_{i\!+\!1} \! - v\, \vec{u}_{i\!+\!1} \, \Delta t - \mu \vec{\nabla} U(\vec{r}_{i\!+\!1}) \Delta t + \sqrt{2D\Delta t} \, \boldsymbol{\xi}_{i\!+\!1} , \\ \label{beom2}
\vartheta_{i} &=& \vartheta_{i\!+\!1} + \sqrt{2D_{\vartheta}\Delta t} \, \eta_{i\!+\!1} \; .
\end{eqnarray}
Note that Eq.~\eqref{beom} is formally equivalent to Eq.~\eqref{eom} but with a flipped sign of the driving velocity, treating this term as odd under time-reversal symmetry~\cite{Shankar2018}.
Furthermore, the sign of the potential gradient term guarantees that in the limit $v \rightarrow 0$ Eq.~(\ref{beom}) reduces to the correct backward dynamics of a passive Brownian particle (see SM for details).
From these equations, it is possible to obtain an analytic expression of the backward-transition probability (see SM for details)
\begin{equation} \label{eq_backwardprob}
\begin{split}
\bar{p} & (w_{i\!+\!1}\! \rightarrow\! w_{i}) = p(w_{i}\! \rightarrow\! w_{i\!+\!1}) \, \frac{\pi(\vec{r}_{i})}{\pi(\vec{r}_{i\!+\!1})} \times \\
&  \dfrac{\exp\Big\{\!-\dfrac{v}{2D}\vec{u}_{i\!+\!1} \cdot \big(\vec{r}_{i}-\vec{r}_{i\!+\!1}+\mu \vec{\nabla} U(\vec{r}_{i\!+\!1}) \Delta t \big)\Big\}}{\exp\Big\{\!-\dfrac{v}{2D}\vec{u}_{i} \cdot \big(\vec{r}_{i}-\vec{r}_{i\!+\!1}-\mu \vec{\nabla} U(\vec{r}_{i}) \Delta t \big)\Big\}} \; ,
\end{split}
\end{equation}
where $\pi(\vec{r}) \propto e^{-\beta U({\vec r})}$ is the Boltzmann distribution for a passive particle.
We stress that the first line of Eq.~\eqref{eq_backwardprob} is the result for a passive Brownian particle, while the second line represents the correction term accounting for the microscopic irreversibility of active Brownian dynamics.
Similar results expressing the ratio between the forward and backward probability may be obtained by a direct time-reversal transformation within the path integral formulation~\cite{fala2016} or by means of Crooks-like relations for entropy production~\cite{croo2011}, in a similar fashion to what has been done in different contexts~\cite{dabe2019,spec2016,chau2014}.

\begin{figure}[t!]
\centering
\includegraphics[width=0.5\textwidth]{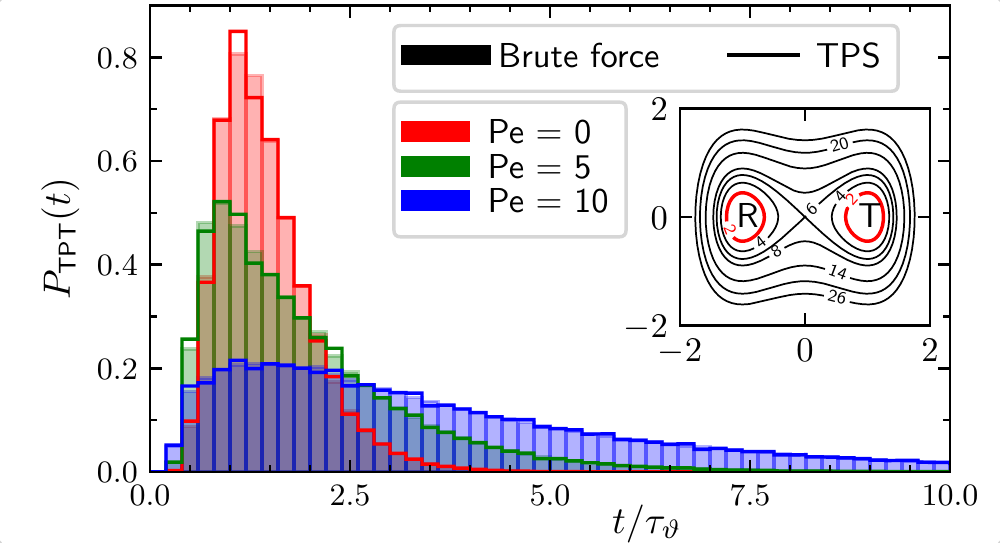}
\vspace{-15pt}
\caption{Distribution of Transition-path times (TPTs) at different P{\'e}clet numbers. 
Each distribution is obtained from $10^6$ different TPTs. ($\tau_{\vartheta} = 1/D_{\vartheta}$)}\label{fig:TPT}
\end{figure}

After combining all terms, the final expression for the acceptance probability can be obtained.
This formula is rather lengthy and can be found in SM.
Here we limit ourselves to noting that in the limit of vanishing activity ($v \to 0$) we recover the standard TPS formula for passive Brownian dynamics.
We also stress that the acceptance probability involves the steady-state distribution of microstates in the reactant basin $\rho(w)$, which is in general not known analytically.
We numerically estimated this distribution from a frequency histogram of the microstates in the reactant basin visited by active Brownian trajectories generated by solving numerically Eq.~\eqref{eom} (see SM).

\begin{figure}[t!]
\centering
\includegraphics[width=0.5\textwidth]{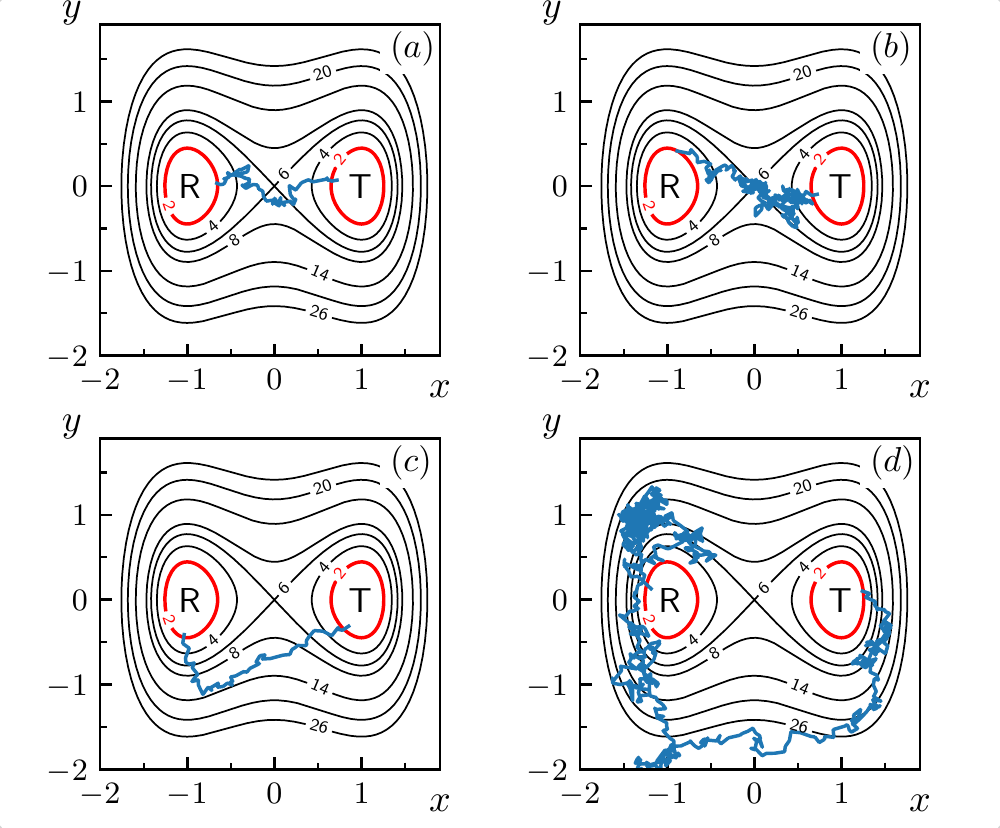}
\vspace{-15pt}
\caption{(a-b) A fast and a slow reactive path ( $t_{\text{TPT}}/\tau_{\vartheta}=0.65$ and $2.37$ respectively) of a passive Brownian particle ($\text{Pe}=0$). The red contour lines are the borders of the R and T basins. (c-d) A fast and a slow reactive path ($t_{\text{TPT}}/\tau_{\vartheta}=0.53$ and $8.36$ respectively) 
of a ABP ($\text{Pe}=10$).  }\label{fig:traj}
\end{figure}

\begin{figure*}[t!]
\centering
\includegraphics[width=1.0\textwidth]{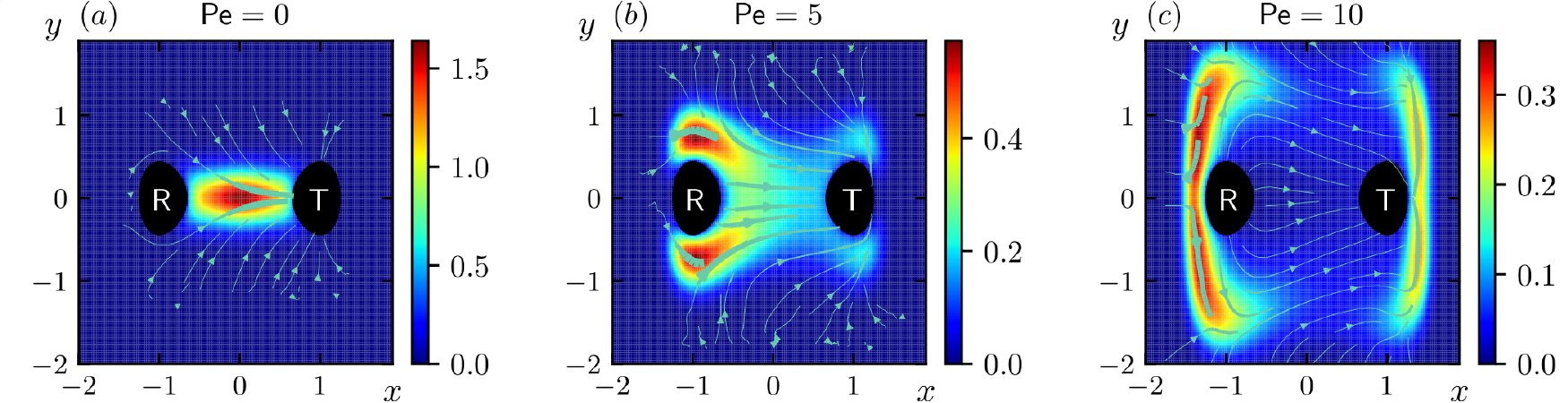}
\vspace{-15pt}
\caption{(a-c) Reactive probability density $m(\vec{r})$ (color map) and 
field lines of the reactive current $\vec{J}(\vec{r})$  (cyan arrows) 
at different P{\'e}clet numbers.
\label{fig:mJ}}
\end{figure*}

We use our TPS algorithm to characterize transition pathways of an ABP reaching a target region crossing the energy barrier in the two-dimensional energy surface 
\begin{equation} \label{eq_potential}
U(x,y)= k_x \, (x^2 - x_0^2)^2 + \dfrac{k_y}{2} y^2 \; ,
\end{equation}
which provides the paradigmatic example of a double well problem. We set $x_0=1$, $k_x=6$ and $k_y=20$ and  measure energy in units of an effective thermal energy $k_B T_{\text{eff}} := D/\mu$.
We define a reactive trajectory as one leaving the reactant region (defined as the set of points with $U(x,y) \le 2 k_BT_{\text{eff}}$ and $x<0$) and reaching the target region (defined by $U(x,y) \le 2 k_BT_{\text{eff}}$ and $x>0$), before returning to the reactant.
Here we study the ABP behavior for three different values of the P{\'e}clet number, $\text{Pe} :=  v \sqrt{3 /4DD_{\vartheta}}$, a dimensionless measure of the activity: $\text{Pe}=0$ ($v=0$), $\text{Pe}=5$ ($v= 1.83$) and $\text{Pe}=10$ ($v=3.65$) while keeping fixed $D=0.1$, $D_{\vartheta}=1$ and $\mu=0.1$.
Important insight on the ABPs behavior in a potential well has already been achieved, for example the escape rates do not follow Kramers theory~\cite{geis2016,roma2012} and they depend explicitly on the full potential barrier~\cite{Woillez2019}.
Here we first consider the distribution of transition-path times (TPTs), i.e. the time durations of reactive trajectories.
This observable has received considerable attention in the context of passive dynamics, both from experimentalists~\cite{chun2009,neup2016,neup2017,neup2018} and theorists~\cite{sega2007,zhan2007,facc2012,kim2015,maka2015,lale2017,bere2017,cara2018,bere2019,cara2020}, because it carries information about the reactive dynamics.
Yet, it appears that TPTs in the presence of activity have been studied only in the case of a one-dimensional particle crossing a parabolic barrier~\cite{carl2018}.

The TPT distribution obtained using our TPS algorithm reproduces well the one obtained by direct integration of the stochastic equations of motion (see Fig.~\ref{fig:TPT}) for the different values of the P{\'e}clet number.
The computational advantage of sampling by TPS relative to brute force simulations is very high at low activities but drops significantly at very large activities.
In the present landscape, the ratio between the time needed to obtain $10^6$ reactive paths with TPS and brute force simulations is $\sim 0.001$, $\sim 0.12$ and $\sim 0.27$ for $\text{Pe}=0$, $\text{Pe}=5$ and $\text{Pe}=10$, respectively.
This decrease in efficiency strictly follows the increase with the activity of crossing rates (see SM) and we expect that the computational advantage of TPS for active particles increases for larger barriers.
Note that, if the TPS scheme is modified accordingly, these results are also confirmed when a different backward dynamics is adopted.
However, this could strongly affect the algorithm's efficiency (see section IV in SM).

Our results show that the average TPT grows with the P{\'e}clet number, while the distribution becomes broader and broader.
To investigate the origin of this difference, it is instructive to analyze the typical transition pathways.
For a passive particle the slow and fast trajectories are quite similar, with the reactive paths narrowly focused around the minimum-energy path crossing the barrier (see Fig.~\ref{fig:traj} a-b).
In contrast, for active particles the trajectories associated with fast and slow transitions are qualitatively very different.
Namely, fast active transition pathways are similar to passive ones (see Fig.~\ref{fig:traj}c).
In contrast, the main contribution to the right tail of the TPT distribution comes from trajectories which leave the reactant basin in the direction opposite to energy saddle point (see Fig.~\ref{fig:traj}d)~\cite{long_TPT}.

This mechanism is confirmed by a systematic statistical analysis of the reactive processes, based on computing the transition-path density, $m(\vec{r})$, and the transition current, $\vec{J}(\vec{r})$~\cite{wein2010}.
Here $m(\vec{r})$ measures the probability that a transition path visits a specific position $\vec{r}$ in the reactive region, while $\vec{J}(\vec{r})$ provides the information on the probability current.
At vanishing $\text{Pe}$ the transition-path density is highest in the saddle point region, while upon adding activity the transition-path density becomes largest in the regions behind the basins, as shown in Fig.~\ref{fig:mJ}.
This behavior is reflected in the reactive current (see Fig.~\ref{fig:mJ}) as well as in the marginalized reactive probability density $m(x) = \int m(x,y) d y$ and $m(y) = \int m(x,y) d x$ (see SM).

Altogether, our results demonstrate that active target search is qualitatively different from that performed by passive systems.
In particular, the observed two order of magnitude increase of the target-finding rates due to the self-propulsion (see SM) is concomitant with the observation that the ABPs reach the target by exploring regions of the energy landscape that are effectively inaccessible for the passive particle.
While one could still naively expect that also for ABPs most of the reactive paths are rather short and develop close to the minimum energy path, we show that long-lasting paths become more and more frequent with increasing activity.
Moreover, in contrast to the passive case, the structure of the transition pathways of active particles qualitatively changes as a function of its TPT: while short trajectories travel along the minimum energy path (in qualitative analogy with the passive case), long-lasting trajectories reach the target in a counter-intuitive way, by climbing the energy surface in the direction opposite to the transition state and then ``surfing'' high energy regions in the potential energy landscape, before landing into the target.
Overall, this picture is confirmed also when considering other energy landscapes and different set of parameters (see SM).
Interestingly, the average TPT increases with the activity of the particle while the opposite behavior is observed in one-dimensional systems~\cite{carl2018}.
Yet, there the reactive pathways reaching the target from the back cannot even occur, correspondingly the observed trend underlines the non-trivial interplay arising between the activity of the particle, the dimensionality of the system, and the environment topology.

In summary, we have addressed the problem of characterizing the structure and kinetics of rare transition pathways undergone by ABPs in search for a target.
To this end, we have derived and validated an extension of TPS which can be used to directly sample rare events undergone by ABPs.
For the first time we have shown that the explicit breaking of microscopic reversibility typical of non-equilibrium systems does not prevent from exploiting backward dynamics in order to develop an efficient TPS algorithm.
Using our TPS algorithm, we have compared the behavior of active and passive Brownian particles reaching a target in a two-dimensional energy landscape characterized by a high energy barrier.

Our results show that, far from equilibrium conditions, significant differences emerge between the reactive kinetics of active and passive particles.
ABPs behave as ``explorers'': once left the reactant basin it may take longer for them to reach the target but they make many more successful attempts and do that by visiting a large portion of the surrounding space.
In contrast, the passive particles are more ``creatures of habit'': they like to stay in their basin and when moving to the target they are very quick and always follow the same path.
We expect similar counter-intuitive target search patterns to be found in a wide range of physical systems of biological, chemical, and technological relevance and the enhanced path sampling scheme developed in this work provides a powerful tool to investigate these processes in a computationally efficient way.

The mathematical scheme we adopted to derive the acceptance rule for ABPs may in principle be applied to a wider class of irreversible systems, as long as it is possible to identify a backward dynamics with a well-defined path probability density and as long as the trajectories obtained by joining the backward and the forward segments remain ``smooth'' (i.e. they look like the trajectories obtained with only forward dynamics).
Straightforward examples include chiral ABPs~\cite{vanTeeffelen2008}, active particles with anisotropic diffusion~\cite{Kurzthaler2016}, and active Ornstein-Uhlenbeck particles~\cite{Bonilla2019,Caprini2019}.
Possibly, one could deal also with particles that orient themselves along spatial gradients~\cite{Pohl2014} by including a bias term in Eq.~(\ref{eom2}) and leaving untouched Eq.~(\ref{eom}-\ref{beom}-\ref{beom2}) but this, as well every other possible choice, should be verified by checking the feasibility and efficiency of the resulting TPS scheme.
Finally, the idea that backward shooting is possible also when there is a lack of microscopic reversibility may have applications going beyond TPS, as conceiving new enhanced sampling algorithms or generalizing existing ones.

\begin{acknowledgments}
We thank Peter G. Bolhuis, Oleksandr Chepizhko, and Hartmut L{\"o}wen for useful discussions. This work was supported by the Austrian Science Fund (FWF): P28687-N27.
\end{acknowledgments}


%

\clearpage
\widetext
\begin{center}
\textbf{\large Supplemental Material for \\ `Target search of active agents crossing high energy barriers'} \\
\medskip

Luigi Zanovello$^{\, 1,2}$, Michele Caraglio$^{\, 1}$, Thomas Franosch$^{\, 1}$, and Pietro Faccioli$^{\, 2, 3}$ \\
\medskip

$1$ \textit{Institut f\"ur Theoretische Physik, Universit\"at Innsbruck, Technikerstra{\ss}e 21A, A-6020, Innsbruck, Austria} \\
$2$ \textit{Dipartimento di Fisica, Universit{\`a} degli studi di Trento, Via Sommarive 14, 38123 Trento, Italy} \\
$3$ \textit{INFN-TIFPA, Via Sommarive 14, 38123 Trento, Italy}
\end{center}

\setcounter{equation}{0}
\setcounter{figure}{0}
\setcounter{table}{0}
\setcounter{page}{1}
\makeatletter
\renewcommand{\theequation}{S\arabic{equation}}
\renewcommand{\thefigure}{S\arabic{figure}}
\renewcommand{\bibnumfmt}[1]{[S#1]}
\renewcommand{\citenumfont}[1]{S#1}

\section{Stochastic path integral for active particles and derivation of the backward probability}

Here we provide the derivation of the forward transition probability in a discrete dynamics as in Eq.~(1-2) of the main text together with its path integral formulation.
We also derive the backward transition probability.

Considering equations of motion of the ABP --Eq.~(1-2) in the main text--, and that the components of the vector noise $\boldsymbol{\xi}_i=(\xi_{x,i},\xi_{y,i})$ and of the scalar noise $\eta_i$ are independent random variables, distributed according to a Gaussian with zero average and unit variance, one can readily write down the probability for a transition between two microstates:
\begin{eqnarray} \label{eq_forwardprob}
p(w_{i}\! \rightarrow\! w_{i\!+\!1}) = 
\mathcal{N} \exp\Big\{\!-\frac{(\vartheta_{i\!+\!1}-\vartheta_{i})^2}{4D_{\vartheta}\Delta t}\Big\}  
\exp\Big\{\!-\frac{\left(\vec{r}_{i\!+\!1}-\vec{r}_{i}-v\vec{u}_{i} \Delta t +\mu \vec{\nabla} U(\vec{r}_{i})  \Delta t \right)^2}{4D\Delta t}\Big\} = \nonumber \\
\mathcal{N}' \exp\Big\{\!-\frac{(\vartheta_{i\!+\!1}-\vartheta_{i})^2}{4D_{\vartheta}\Delta t}\Big\}
\exp\Big\{\!-\frac{v^2 \Delta t}{4D}\Big\}
\exp\Big\{\!-\frac{v}{2D} \vec{u}_{i}   \cdot  \left(\vec{r}_{i}-\vec{r}_{i\!+\!1}-\mu \vec{\nabla} U(\vec{r}_{i} ) \Delta t \right)\Big\} \pi(\vec{r}_{i}\! \rightarrow\! \vec{r}_{i\!+\!1}) \; ,
\end{eqnarray}
where $\mathcal{N}$ and $\mathcal{N}'$ are normalization constants and in the second line we have factorized the probability $\pi(\vec{r}_{i}\! \rightarrow\! \vec{r}_{i\!+\!1})$ for a passive particle going from $\vec{r}_{i}$ to $\vec{r}_{i\!+\!1}$ in a time step $\Delta t$.

Considering that the succession of the states $w_i=(\vec{r}_i,\vartheta_i)$ constitutes a Markov process, the probability of finding the system in a specific microstate $w_{N}$ at time $\tau = N \Delta t$ provided the initial condition $w_{0}$ at time $t=0$ can be obtained as
\begin{equation}
\mathcal{P}(w_{N}|w_{0}) = \prod_{i=0}^{N-1} p(w_{i}\! \rightarrow\! w_{i\!+\!1}) \; .
\end{equation}
The latter, in the limit of $\Delta t \rightarrow 0$, can be rewritten using the following stochastic path integral formulation:
\begin{equation}
\mathcal{P}(w_{N}|w_{0}) = \mathcal{Z}^{-1} \int_{\vec{r}_0}^{\vec{r}_N}\!\mathcal{D} \vec{r}\!\int\!\mathcal{D} \vartheta \,\exp{\left(-\frac{1}{4D_{\vartheta}}S_{\text{rot}}[\vartheta]\right)}  \exp{\left(-\frac{1}{4D}S_{\text{trans}}[\vec{r},\vartheta]\right)} \; ,
\end{equation}
where $\mathcal{Z}$ is again a normalization constant and $S_{\text{rot}}$ and $S_{\text{trans}}$
are functionals encoding the rotational and translational noise:
\begin{equation}
S_{\text{rot}}[\vartheta] = \int_{0}^{\tau}\! d t \, \left[ \dot{\vartheta}(t)\right]^{2} \; ,
\end{equation}
\begin{equation}
S_{\text{trans}}[\vec{r},\vec{u}] = \int_{0}^{\tau}\! d t \,  \Big[\dot{\vec{r}}(t) -v\, \vec{u}(t) + \mu \vec{\nabla} U\big(\vec{r}(t)\big)\Big]^{2}  \; .
\end{equation}
Note that $S_{\text{trans}}$ is the active equivalent of the Onsager--Machlup functional of a passive particle~\cite{onsa1953,mach1953}.

Considering the discrete backward dynamics defined by Eq. (7-8) of the main text and using the same argument that led us to Eq.~(\ref{eq_forwardprob}), the backward transition probability reads
\begin{eqnarray} \label{eq_backwardprob}
\bar{p} (w_{i\!+\!1}\! \rightarrow\! w_{i}) = 
\mathcal{N} \exp\Big\{\!-\frac{(\vartheta_{i}-\vartheta_{i\!+\!1})^2}{4D_{\vartheta}\Delta t}\Big\}  
\exp\Big\{\!-\frac{\left(\vec{r}_{i}-\vec{r}_{i\!+\!1}+v\vec{u}_{i\!+\!1} \Delta t + \mu \vec{\nabla} U(\vec{r}_{i\!+\!1})  \Delta t \right)^2}{4D\Delta t}\Big\} = \nonumber \\
\mathcal{N}' \exp\Big\{\!-\frac{(\vartheta_{i}-\vartheta_{i\!+\!1})^2}{4D_{\vartheta}\Delta t}\Big\}
\exp\Big\{\!-\frac{v^2 \Delta t}{4D}\Big\} 
\exp\Big\{\!-\frac{v}{2D}\vec{u}_{i\!+\!1}   \cdot  \left(\vec{r}_{i}-\vec{r}_{i\!+\!1}+\mu \vec{\nabla} U(\vec{r}_{i\!+\!1} ) \Delta t \right)\Big\} \pi(\vec{r}_{i\!+\!1}\! \rightarrow\! \vec{r}_{i}) \; ,
\end{eqnarray}
which easily leads to Eq. (9) of the main text once one considers that detailed balance condition
\begin{equation} \label{eq_detbalance}
 \pi(\vec{r}_{i\!+\!1}) \pi(\vec{r}_{i\!+\!1}\! \rightarrow\! \vec{r}_{i}) = \pi(\vec{r}_{i}) \pi(\vec{r}_{i}\! \rightarrow\! \vec{r}_{i\!+\!1}) ,
\end{equation}
holds for a passive particle.

\section{Final acceptance probability}

Here we present the final form of the acceptance probability implemented in our generalized TPS algorithm obtained by combining Eq.~(3-6,9) of the main paper.

\begin{figure}[h!]
\centering
\includegraphics[width=\textwidth]{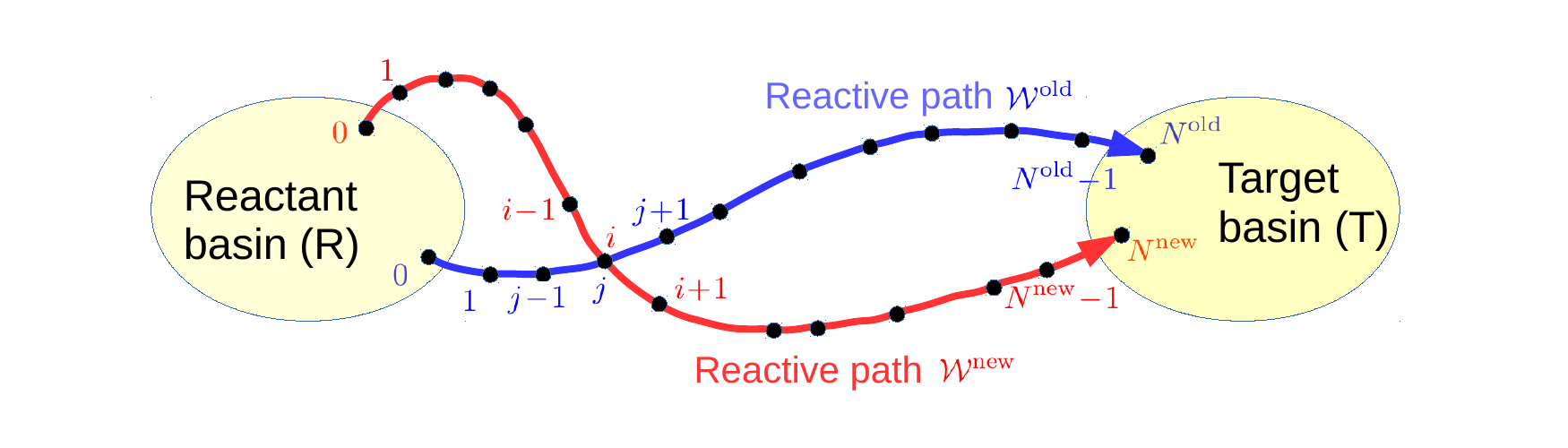}
\vspace{-25pt}
\caption{Sketch representing the TPS algorithm. A new reactive trajectory $\mathcal{W}^{\text{new}}$ is generated by solving the system's  equations of motion forward and backward in time, starting from $w_j^{\text{new}} = w_i^{\text{old}}$, with $w_i^{\text{old}}$ a frame of the old reactive path $\mathcal{W}^{\text{old}}$.}\label{fig:sketch}
\end{figure}

We recall that a newly attempted transition path $\mathcal{W}^{\text{new}}$ generated starting from an old path $\mathcal{W}^{\text{old}}$ (see Fig.~\ref{fig:sketch}) is proposed according to a three-step procedure: 
First, a microstate $w_i^{\text{old}}$ is randomly picked from the frames in $\mathcal{W}^{\text{old}}$.
Next, we select as shooting point $w_j^{\text{new}} = w_i^{\text{old}}$.
Finally, a new trial trajectory $\mathcal{W}^{\text{new}}$ is obtained by solving the system's  equations of motion forward and backward in time, starting from $w_j^{\text{new}}$.
The resulting new trajectory $\mathcal{W}^{\text{new}}$ is then accepted with the following probability

\begin{equation} \label{eq_accprobfinal}
\begin{split}
& \mathcal{P}_{\text{acc}}  \left[\mathcal{W}^{\text{old}} \!\rightarrow\! \mathcal{W}^{\text{new}}\right]  = h \left[  \mathcal{W}^{\text{new}} \right] \, 
\min \Bigg\{1,  \dfrac{N^{\text{old}}}{N^{\text{new}}} \dfrac{\rho(w^{\text{new}}_0)}{\rho(w^{\text{old}}_0)}
\frac{\pi (\vec{r}_{0}^{\text{ old}})}{\pi(\vec{r}_{0}^{\text{ new}})} \times \\
& \quad  \prod_{k=0}^{j-1} \exp \Big\{\! -\dfrac{v}{2D}  \vec{u}_{k\!+\!1}^{\text{ old}} \cdot \big(\vec{r}_{k}^{\text{ old}}-\vec{r}_{k\!+\!1}^{\text{ old}}+\mu \vec{\nabla} U(\vec{r}_{k\!+\!1}^{\text{ old}}) \Delta t \big)  \Big\} \times \\
& \quad  \prod_{k=0}^{j-1} \exp \Big\{ \dfrac{v}{2D}  \vec{u}_{k}^{\text{ old}} \cdot \big(\vec{r}_{k}^{\text{ old}}-\vec{r}_{k\!+\!1}^{\text{ old}}-\mu \vec{\nabla} U(\vec{r}_{k}^{\text{ old}}) \Delta t \big)  \Big\} \times \\
& \quad \prod_{k=0}^{i-1} \exp \Big\{ \dfrac{v}{2D}  \vec{u}_{k\!+\!1}^{\text{ new}} \cdot \big(\vec{r}_{k}^{\text{ new}}-\vec{r}_{k\!+\!1}^{\text{ new}}+\mu \vec{\nabla} U(\vec{r}_{k\!+\!1}^{\text{ new}}) \Delta t \big)  \Big\} \times \\
& \quad  \prod_{k=0}^{i-1} \exp \Big\{\! -\dfrac{v}{2D}  \vec{u}_{k}^{\text{ new}} \cdot \big(\vec{r}_{k}^{\text{ new}}-\vec{r}_{k\!+\!1}^{\text{ new}}-\mu \vec{\nabla} U(\vec{r}_{k}^{\text{ new}}) \Delta t \big)  \Big\} \Bigg\} \; .
\end{split}
\end{equation}

Note that in the limit of vanishing activity ($v \to 0$) the previous equation reduces to the standard TPS acceptance probability for passive Brownian dynamics: $ \mathcal{P}_{\text{acc}}  \left[\mathcal{W}^{\text{old}} \!\rightarrow\! \mathcal{W}^{\text{new}}\right] =  h\left[\mathcal{W}^{\text{new}}\right] \, \min \left\{1, N^{\text{old}}/N^{\text{new}}\right\}$.

\section{Steady state distribution within the reactant basin}

\begin{figure}
\centering
\includegraphics[width=\textwidth]{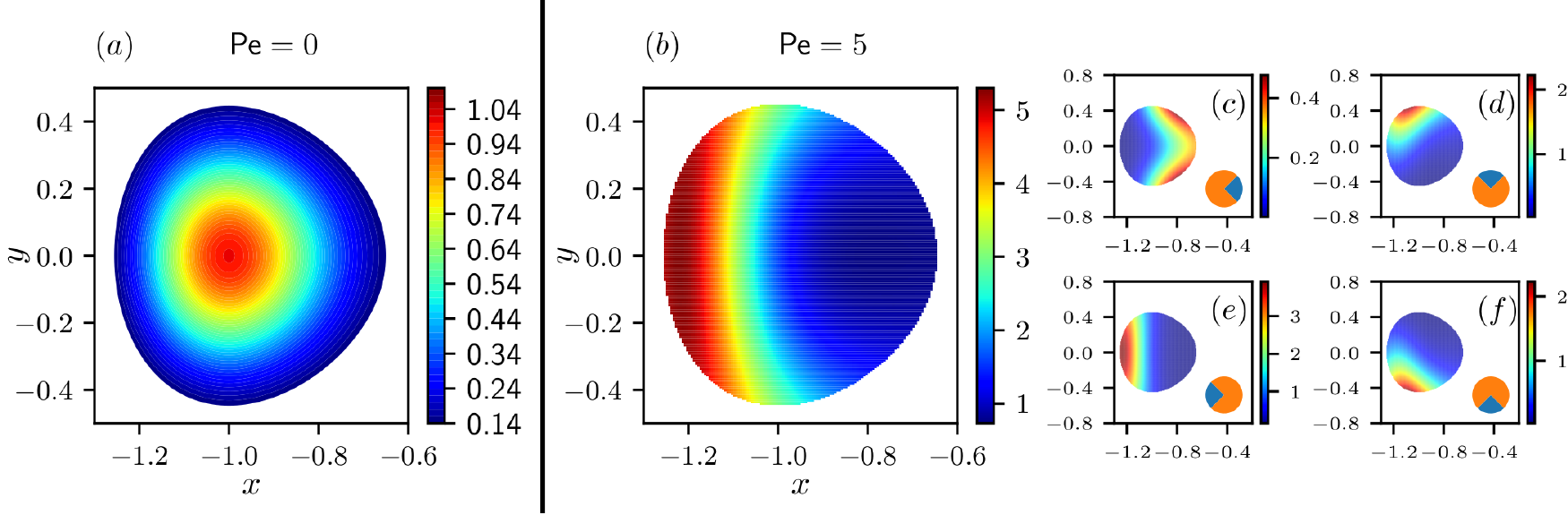}
\caption{Steady state distributions within the reactant basin. 
(a) equilibrium distribution for a passive particle, with $k_{x} = 6$ and $k_y=20$.
(b) steady state distribution for an active particle with $\text{Pe} = 5$, $k_{x} = 6$ and $k_y=20$. For each $(x,y)$ point, the distribution value is obtained as the cumulative distribution over all the possible angles $\vartheta$ of the self-propulsion speed.
(c-f) steady state distribution marginalized on the value of the angle $\vartheta$. The four plot report distributions marginalized on the four quadrant: $-\pi/4 < \vartheta \leqslant \pi/4$ (c), $\pi/4 < \vartheta \leqslant 3\pi/4$ (d), $3\pi/4 < \vartheta \leqslant 5\pi/4$ (e) and  $5\pi/4 < \vartheta \leqslant 7\pi/4$ (f). See also the blue portion in the pie plots.
}\label{fig:rho}
\end{figure}

To implement the TPS algorithm the steady state distribution within the reactant basin is needed. 
The passive particle is Boltzmann-distributed (see Fig.~\ref{fig:rho}a).
In the case of ABP, the steady state distribution is not Boltzmann-distributed and is obtained from a frequency histogram of the microstates in the reactant basin visited by active Brownian trajectories generated by solving numerically Eq.~(1-2) of the main text.
The resulting distribution shows that when the ABP is inside the reactants basin, it is more likely to be on the left side of the basin (see Fig.~\ref{fig:rho}b). 
Furthermore, the positional probability distribution of an ABP inside the basin strongly depends on the angle $\vartheta$ determining the direction of its velocity (see Fig.~\ref{fig:rho}c-f)
The results show that the ABP is more likely to have a velocity that points in the opposite direction with respect to the center of the basin.
The observed overall behavior is due to the fact that the left side of the basin is characterized by a steeper potential wall.
When the ABP has a velocity pointing towards positive $x$ values, it can easily exit the reactant basin.
In contrast, when the ABP has a velocity pointing towards the steeper potential wall (direction towards negative $x$ values), it is more likely that the particle remains inside the basin, at least for a time of the order $\tau_{\vartheta}$, which is the typical time necessary for the particle to change the velocity direction.

\section{Definition of the backward dynamics}

Here we collect a few remarks about our suggested rule for the backward shooting:
\begin{eqnarray}\label{beom}
\vec{r}_{i} &=& \vec{r}_{i\!+\!1} \! - v\, \vec{u}_{i\!+\!1} \, \Delta t - \mu \vec{\nabla} U(\vec{r}_{i\!+\!1}) \Delta t + \sqrt{2D\Delta t} \, \boldsymbol{\xi}_{i\!+\!1} , \\ \label{beom2}
\vartheta_{i} &=& \vartheta_{i\!+\!1} + \sqrt{2D_{\vartheta}\Delta t} \, \eta_{i\!+\!1} \; ,
\end{eqnarray}
corresponding to Eq.~(7-8) of the main text.

\begin{figure}
\centering
\includegraphics[width=0.5\textwidth]{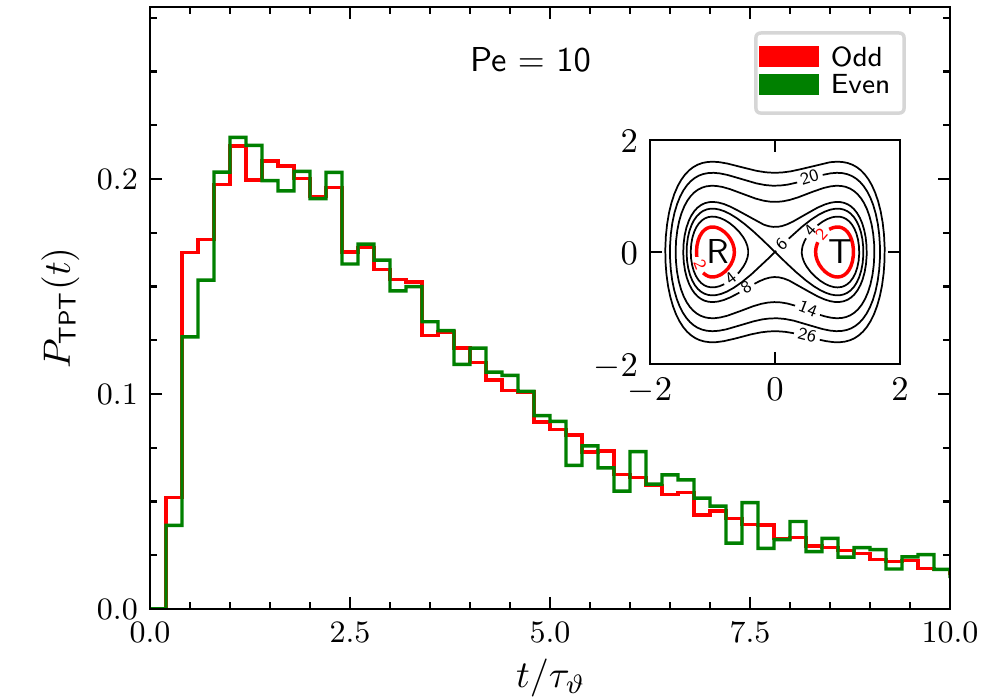}
\vspace{-10pt}
\caption{Comparison of TPT distributions obtained through a TPS scheme in which either odd or even signature of the self-propulsion term in the backward integration step is used (ABP parameters: $D=0.1$, $v=3.65$, $\mu=0.1$, $D_{\vartheta}=1$. Potential parameters: $k_x=6$, $k_y=20$). \label{fig_even_odd}}
\end{figure}

First, we discuss the sign of the self-propulsion term (second term on right side of Eq.~(\ref{beom})).
In order to have trial reactive path (obtained by joining together the backward and forward shoots)  resembling  as much as possible the reactive paths obtained by integrating the equations of motion only forward in time, the self-propulsion term is treated as odd under time reversal symmetry, i.e. at par with particle momentum, it reverses sign under time reversal.
This differs from the even signature in which self propulsion is treated as a non-conservative force and does not change sign under time reversal.
See Ref.~\cite{Shankar2018} for a thorough discussion of the various implication of the choice between the two conventions.
However, one is in principle allowed to adopt an even signature convention for the self-propulsion term, provided that Eq.~(\ref{eq_backwardprob}) and Eq.~(\ref{eq_accprobfinal}) are  modified accordingly. 
Yet, with this modified scheme it is possible to properly sample the reactive path space only in the presence of a non-negligible contribution of the translational diffusion.
This is actually the case for the parameter used to present the results of our manuscript, at the price of  an about $10$-fold decrease in  efficiency.
Fig.~\ref{fig_even_odd} shows that both  schemes using odd signature and  even signature reproduce the same TPT distribution. 
\medskip

Second, we discuss the sign of the conservative force term (third term on right side of Eq.~(\ref{beom}).
In this case we opted for a negative sign for the two following reasons:
\begin{enumerate}
\item As for the self propulsion term, one is in principle allowed to flip the sign of the force term in the backward dynamics.
However, if so the TPS algorithm becomes very inefficient in the sense that new trajectories in the reactive path ensemble are accepted with a very low rate by the Metropolis procedure. 
Indeed, considering this ``flipped sign backward dynamics'' and the model parameters reported in the main text, the acceptance rate of new reactive trajectories turns out to be about 4-6 times smaller.
\item Our algorithm has the benefit of reducing to the standard equilibrium one for vanishing P{\'e}clet number.
In the case of passive particles in the overdamped limit the forward and backward probability are indistinguishable (see ref. [22]).
In fact, the forward and backward probability are related by microscopic reversibility (see ref. [22]):
\begin{equation}
\pi(\vec{r}_{i})
 \pi(\vec{r}_{i}\! \rightarrow\! \vec{r}_{i\!+\!1})   = \pi(\vec{r}_{i\!+\!1}) \bar{\pi}(\vec{r}_{i\!+\!1}\! \rightarrow\! \vec{r}_{i})  \; ,
\end{equation}
and since detailed balance, Eq.~(\ref{eq_detbalance}), also holds, it follows that the backward probability is equal to the forward probability
\begin{equation}
\bar{\pi}(\vec{r}_{i\!+\!1}\! \rightarrow\! \vec{r}_{i}) = \pi(\vec{r}_{i\!+\!1}\! \rightarrow\! \vec{r}_{i}) \; ,
\end{equation}
and that the backward trajectory can be generated using the forward propagation rule.
\end{enumerate}

\section{Detailed analysis for $\text{Pe} = 5$  and $\text{Pe} = 10$}

\begin{figure*}
\centering
\includegraphics[width=0.96\textwidth]{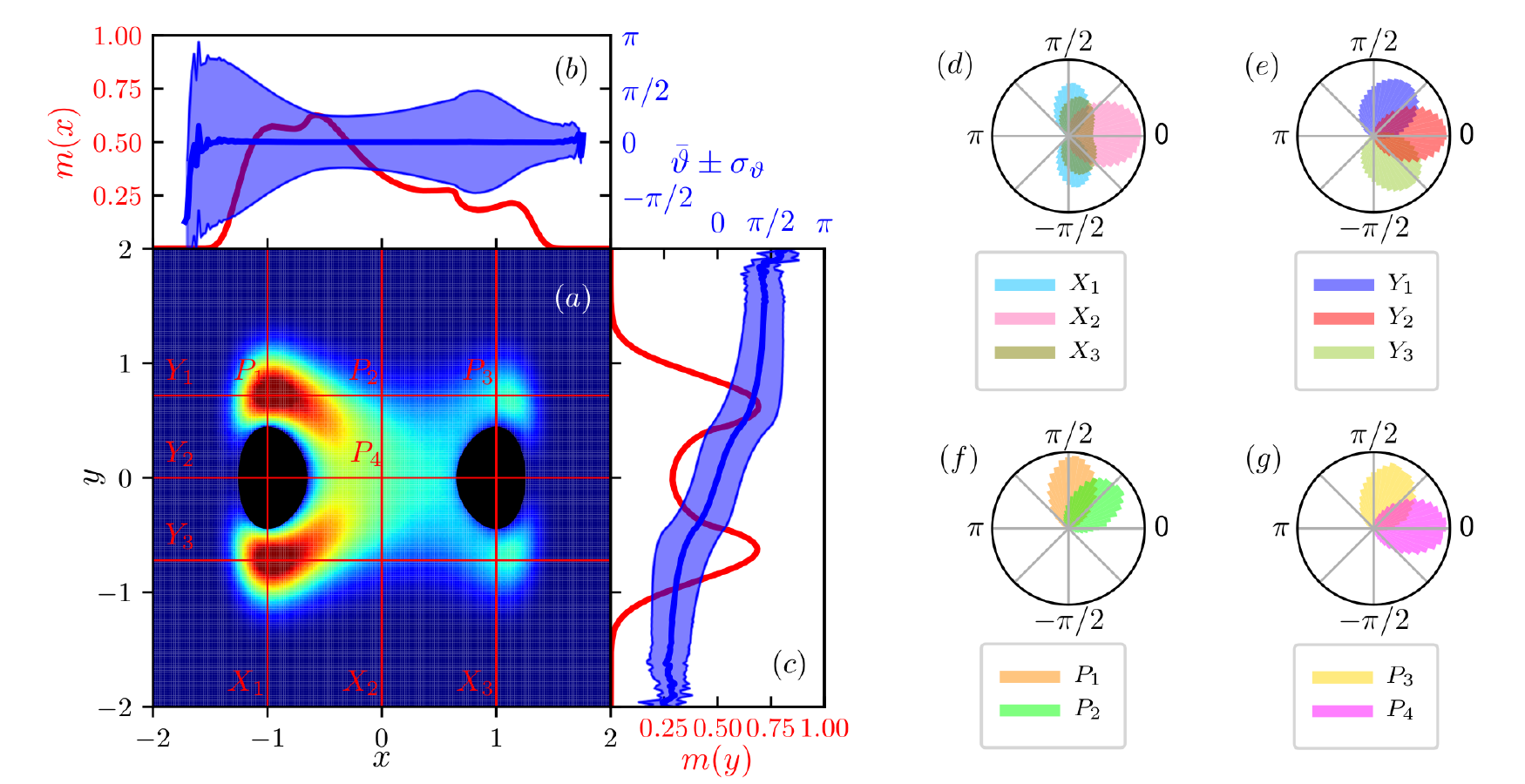}
\vspace{-10pt}
\caption{(a) Reactive probability density $m(\vec{r})$ for $\text{Pe} = 5, k_{x} = 6, k_{y} = 20$. 
(b-c) marginal reactive probability density $m(x)$ ($m(y)$) (red line) and average angle $\bar{\vartheta}$ with its standard deviation $\sigma_{\vartheta}$ (thick blue line) as a function of $x$ ($y$) alone.
(d-e) distribution of angle $\vartheta$ among the states having x-coordinate in the intervals $X_1:[-1-\Delta,-1+\Delta]$, $X_2:[-\Delta,\Delta]$ and $X_3:[1-\Delta,1+\Delta]$ ($Y_1:[-0.72-\Delta,-0.72+\Delta]$, $Y_2:[-\Delta,\Delta]$ and $Y_3:[0.72-\Delta,0.72+\Delta]$) respectively ($\Delta=0.04$).
(f-g) distribution of angle $\vartheta$ in the states $P_1,\ldots,P_4$, at the intersection of previously defined intervals.
Each angular distribution is normalized.}\label{fig:angles}
\end{figure*}

\begin{figure}[h!]
\centering
\includegraphics[width=0.96\textwidth]{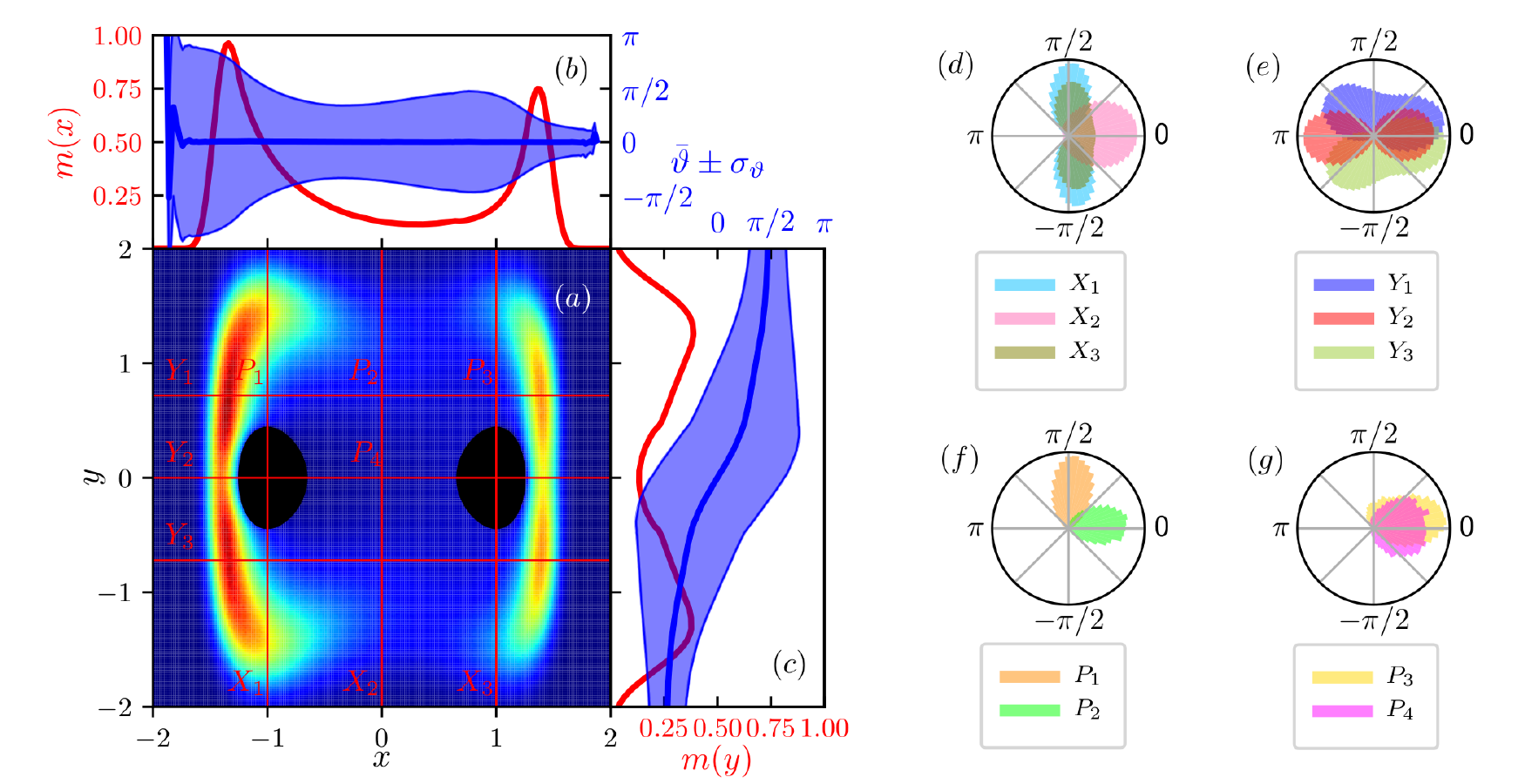}
\vspace{-10pt}
\caption{(a) Reactive probability density $m(\vec{r})$ for $\text{Pe} = 10, k_{x} = 6, k_{y} = 20$. (b-c) marginal reactive probability density $m(x)$ ($m(y)$) (red line) and average angle $\bar{\vartheta}$ with its standard deviation $\sigma_{\vartheta}$ (thick blue line) as a function of $x$ ($y$) alone.
(d-e) distribution of angle $\vartheta$ among the states having x-coordinate in the intervals $X_1:[-1-\Delta,-1+\Delta]$, $X_2:[-\Delta,\Delta]$ and $X_3:[1-\Delta,1+\Delta]$ ($Y_1:[-0.72-\Delta,-0.72+\Delta]$, $Y_2:[-\Delta,\Delta]$ and $Y_3:[0.72-\Delta,0.72+\Delta]$) respectively ($\Delta=0.04$).
(f-g) distribution of angle $\vartheta$ in the states $P_1,.., P_4$ at the intersection of previously defined intervals.
Each angular distribution is normalized.}\label{fig:angular}
\end{figure}

Greater details on the reactive path ensemble can be collected by studying the behavior of the angle $\vartheta$ in the states belonging to the transition paths.
Here we discuss the case of medium activity $\text{Pe}=5$.
Due to symmetry with respect to the axis $y=0$, the average angle as a function of $x$ is zero.
More interestingly, the standard deviation of the angles as a function of $x$ shows that the angles are more focused in forward direction ($\vartheta=0$) just at the right of the reactants basin (see blue lines Fig.~\ref{fig:angles}b).
In fact, the angular distributions of $\vartheta$ conditioned to specific values of $x$ are spread around $\vartheta=0$ at $x=0$, while they are bimodal in the proximity of the basins (see Fig.~\ref{fig:angles}d).
Along the $y$ coordinate, instead, on average the angle points upward for positive $y$ values and downward for negative $y$ values while the standard deviation remains about constant (see Fig.~\ref{fig:angles}c).
This behavior is reflected in the angular distributions of $\vartheta$ conditioned to specific values of $y$, as reported in Fig.~\ref{fig:angles}e.
Angular distributions conditioned to a specific position show that for points close to the reactant basin the most likely velocity angle points in the direction opposite to the basin itself.
However, the angular distribution conditioned to a specific position just above the target basin  displays again angles contained in the first quadrant (see Fig.~\ref{fig:angles} f-g).
Altogether, the picture emerging from Fig.~\ref{fig:angles} is that there are few fast reactive paths similar to the one reported in Fig.~2(c) and many paths similar to the one in Fig.~2(d) (in the main text) that surf along the energy walls before falling in the T basin (see also Fig.~S5).
Fig.~\ref{fig:angular} is equivalent to Fig.~\ref{fig:angles} but in the case of $\text{Pe} = 10$.

\begin{figure}[h!]
\centering
\includegraphics[width=0.7\textwidth]{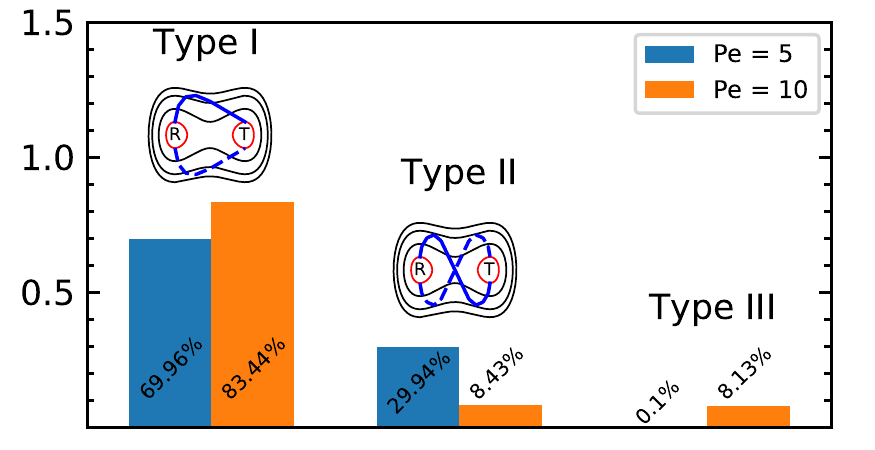}
\vspace{-10pt}
\caption{Frequency of different trajectory types for two values of the P{\'e}clet number, $\text{Pe} = 5$ and $\text{Pe}=10$. A sketch of the trajectory type is given above the bars.}\label{fig:histogram}
\end{figure}

Different transition pathway are sorted in the three categories by studying their behavior in the region $X: -1 < x <1$, when going from the R basin to the T basin: Type I) paths whose positions contained in the region $X$ have only positive (or only negative) $y$ values; Type II) paths whose portion contained in the region $X$ has positive (negative) $y$ coordinates when close R and negative (positive) $y$ coordinates when close to T; Type III) other types of trajectories.
Fig.~\ref{fig:histogram} reports the frequencies with which different reactactive trajectory types are observed.
For $\text{Pe} = 5$ most of the trajectories are of type I, a relevant part is of type II while almost no other types of trajectories are observed. For $\text{Pe} = 10$, instead, the percentage of trajectories of the first kind increases as well as other possible types of trajectories, while the second type of trajectories decreases.
The higher (lower) fraction of transition paths belonging to type I (II) for $\text{Pe} = 10$, in comparison to $\text{Pe} = 5$, is due to the fact that particles with higher activity are more likely to ``surf'' along the potential energy walls.

\section{Dependence on the energy landscape}\label{sec_differentU}

In this section we provide additional information on the target search patterns displayed by ABP in different energy landscapes.
\medskip

\begin{figure}[h!]
\centering
\includegraphics[width=\textwidth]{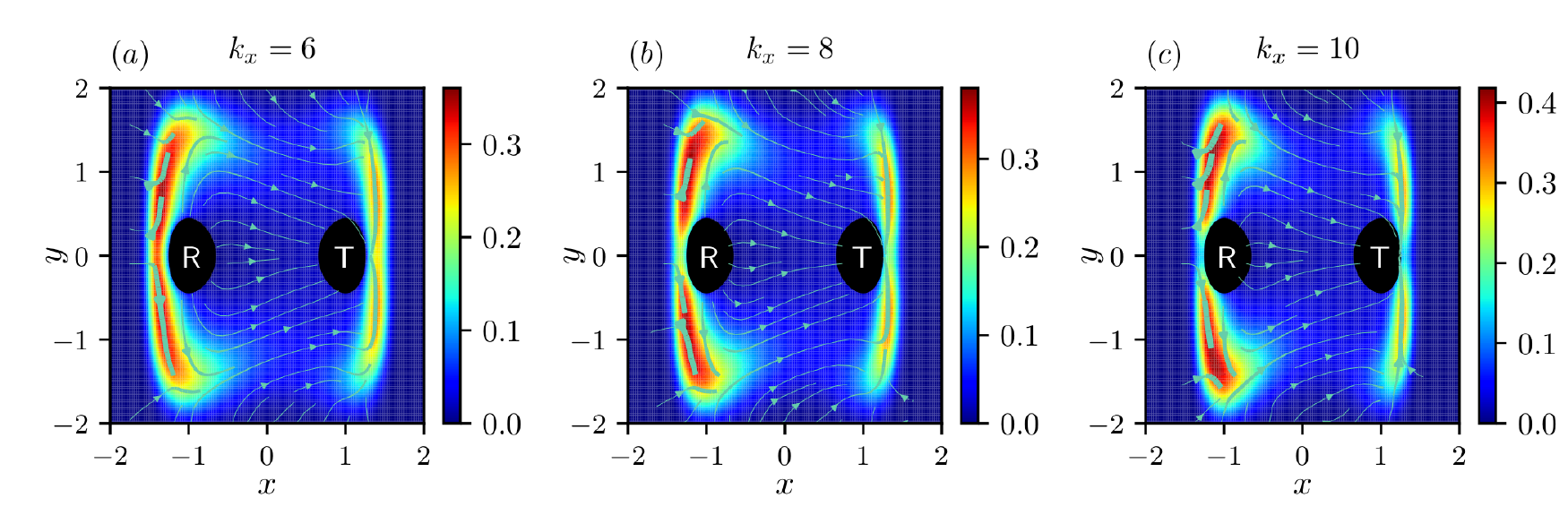}
\caption{(a-c) $m(\vec{r})$ and $\vec{J}(\vec{r})$ in the reactive region for $\text{Pe} = 10$ at different potential stiffnesses $k_{x}$ with $k_y=20$ (ABP parameters: $D=0.1$, $D_{\vartheta}=1$, $\mu=0.1$, $v=3.65$).}\label{fig:deltau}
\end{figure}

First, we consider the same form of energy landscape discussed in the main text but with different potential stiffnesses.
Fig.~\ref{fig:deltau} shows the the transition path density, $m(\vec{r})$, and the transition current, $\vec{J}(\vec{r})$, for $\text{Pe} = 10$ when changing the stiffness parameter $k_x$.
The variation in the quartic part of the potential is not introducing substantial changes in the reactive probability densities and currents, rather they remain quite similar for all explored values of $k_{x}$. The most notable difference resides again in the region explored by the active particles, that is getting closer to the basins along the $x$ direction upon increasing the potential stiffness $k_x$.

\medskip

\begin{figure}[h!]
\centering
\includegraphics[width=1.0\textwidth]{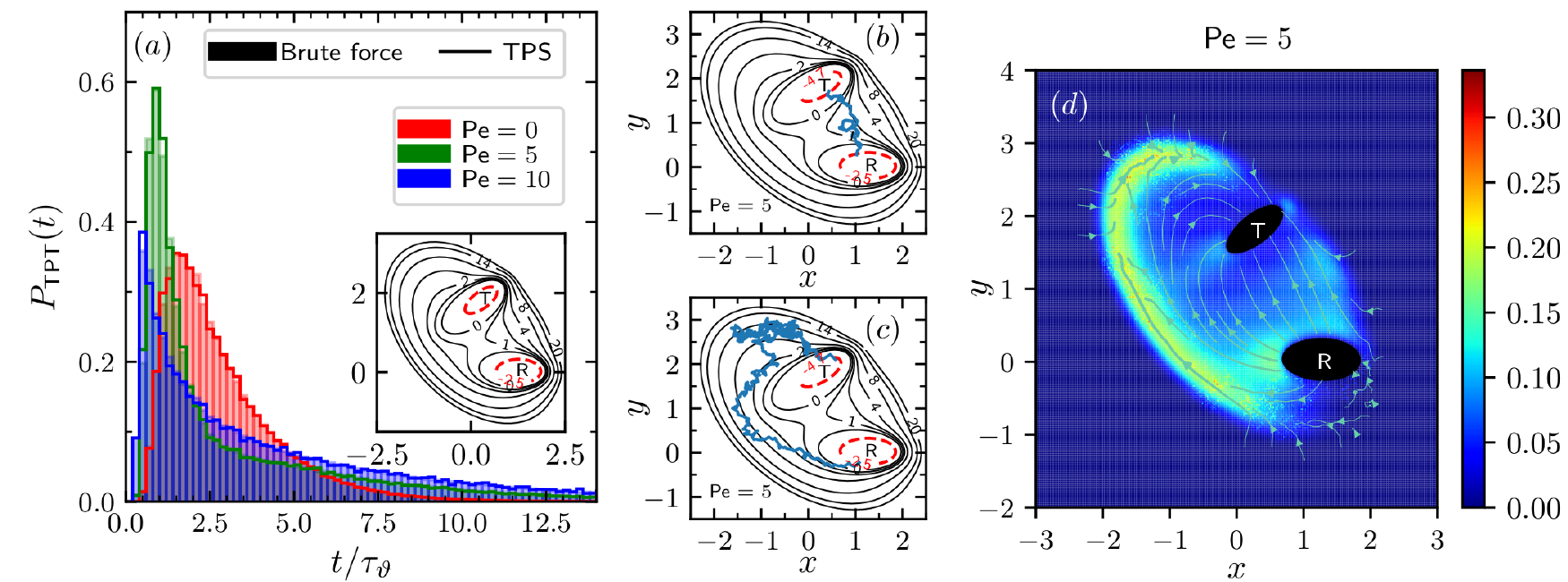}
\caption{(a) Distribution of TPTs at different P{\'e}clet numbers in the energy landscape reported in the inset. Each distribution is obtained from $10^6$ different TPTs.
(b) typical fast reactive paths at $\text{Pe}=5$.
(c) typical long reactive paths at $\text{Pe}=5$.
(d) Reactive probability density $m(\vec{r})$ (color map) and field lines of the reactive current $\vec{J}(\vec{r})$  (cyan arrows) at $\text{Pe}=5$.}\label{fig:differentU}
\end{figure}

Then, we study the target-search scenario in a completely different energy landscape. We propose a M\"{u}ller-like potential of the form:
\begin{equation}
U(x,y) = \sum_{i=1}^{3} K_{i} e^{[a_{i}(x-x_{0,i})^{2}+b_{i}(x-x_{0,i})(y-y_{0,i})+c_{i}(y-y_{0,i})^{2}]} \; ,
\end{equation}
with parameters
\begin{center}
\begin{tabular}{|c|c|c|c|c|c|c|}
\hline 
$i$ & $K_i$ & $a_i$ & $b_i$ & $c_i$ & $x_0^i$ & $y_0^i$ \\ 
\hline 
$1$ & $-10$ & $-0.8$ & $0$ & $-5$ & $1.7$ & $0$ \\ 
$2$ & $-10$ & $-3$ & $6$ & $-5$ & $0.5$ & $2$ \\ 
$3$ & $1$ & $0.7$ & $0.6$ & $0.7$ & $0$ & $1$ \\ 
\hline 
\end{tabular} 
\end{center}
Given this energy landscape, we define the reactant basin (R) as the region with energy lower than $-2.5 k_BT_{\text{eff}}$ around the minimum close to $(x_{0,1},y_{0,1})$ (see inset in Fig.~\ref{fig:differentU}a).
R is about $3.4 k_BT_{\text{eff}}$ deep. 
The target basin is defined as the region with energy lower than $-4.7 k_BT_{\text{eff}}$ around the minimum close to $(x_{0,2},y_{0,2})$. 
T is about $2.7 k_BT_{\text{eff}}$ deep.
The saddle point separating R from T corresponds to an energy of about $0.9 k_B T_{\text{eff}}$.
In contrast to the energy landscape proposed in the main text, this one is highly asymmetric with a curved minumum energy path from R to T.
Also in this case, the TPT distribution obtained using our TPS algorithm reproduces well the one obtained by direct integration of the stochastic equations of motion for all  different considered values of the P{\'e}clet number (see Fig.~\ref{fig:differentU}a).
Furthermore, increasing the activity the mode (i.e. the most likely transition path time) of the distribution shifts to shorter times while its right tail becomes longer, meaning that there are more very fast transition paths (see Fig.~\ref{fig:differentU}b) and, at the same time, an increase in the number of transition paths that take longer detour ``surfing'' high-energy regions before landing in the target region (see Fig.~\ref{fig:differentU}c).
This mechanism is confirmed by an analysis of the transition-path density and of the transition current (see Fig.~\ref{fig:differentU}d for the case $\text{Pe}=5$).

To conclude this section, we note that for the ABPs, the proposed TPS scheme can deal not only with the proposed paradigmatic potential energy surface but with virtually any landscape as long as well-defined reactant and target regions exist.
In particular, in complex energy landscapes, one might be interested in how many typical reaction channels exist and on which transition states or metastable states are visited by the system on its way to the target.
Practically, it is precisely the presence of many local minima that may pose severe issues to the efficiency of the TPS algorithm.
To solve this problem classic TPS was complemented with a quenching technique~\cite{Dellago1998} and the same method can be applied and tested in our case.

\section{Dependence on the rotational diffusion parameter}

\begin{figure}[h!]
\centering
\includegraphics[width=1.0\textwidth]{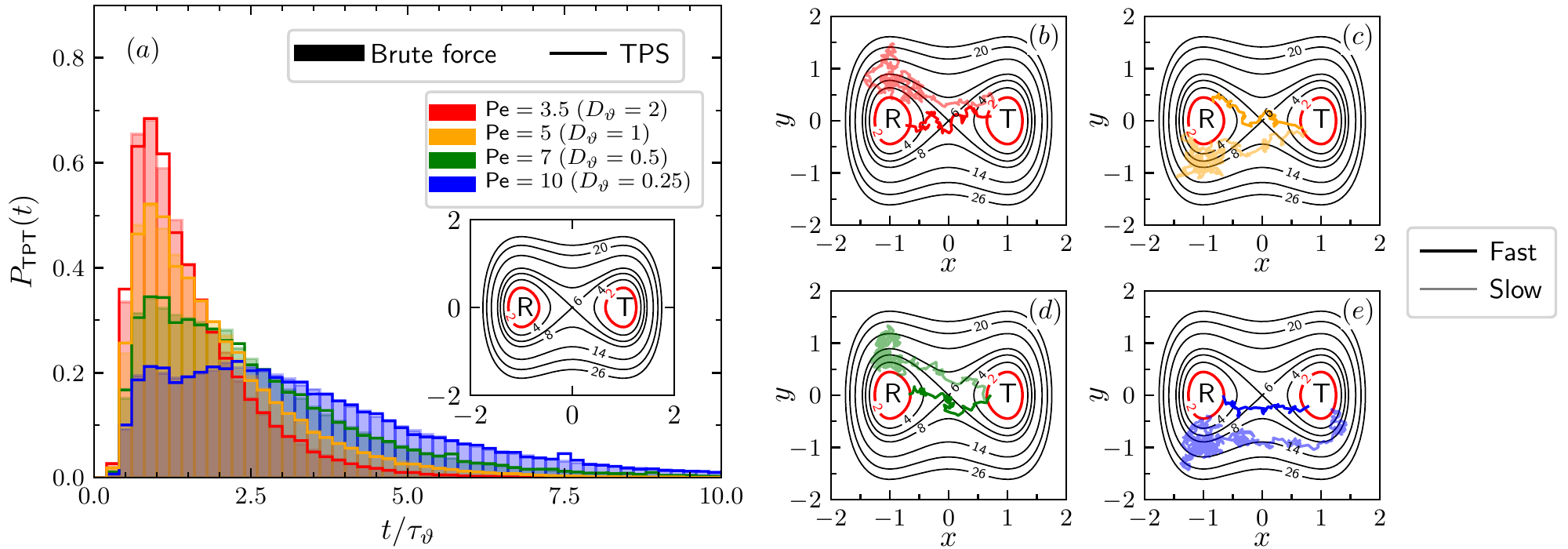}
\caption{(a) Distribution of Transition-path times (TPTs) at different P\'eclet numbers obtained by changing the rotational diffusion coefficient. Each distribution is obtained from $10^6$
different TPTs. (b-e) A fast and a slow reactive path for each P\'eclet number considered in panel a. Color code as in panel a. (ABP parameters: $D=0.1$, $\mu=0.1$, $v=1.83$. Potential parameters: $k_x=6$, $k_y=20$).}\label{fig:Distribution_TPT_vs_Dr}
\end{figure}

\begin{figure}[h!]
\centering
\includegraphics[width=\textwidth]{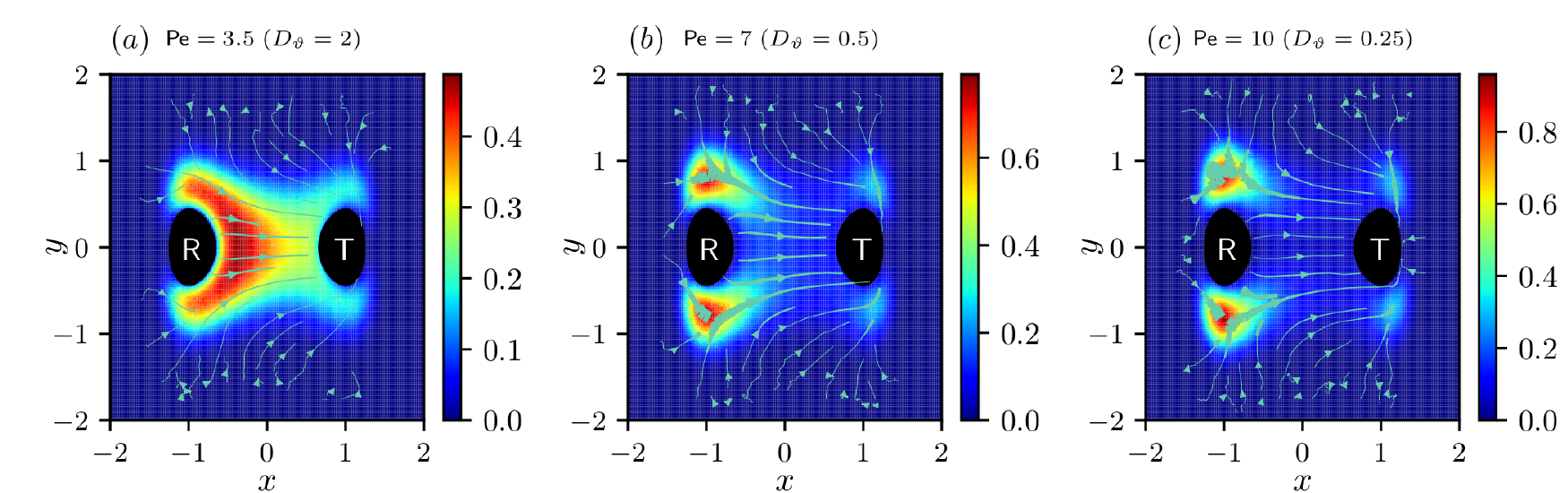}
\caption{(a-c) $m(\vec{r})$ and $\vec{J}(\vec{r})$ in the reactive region for different P{\'e}clet number obtained by changing $D_{\vartheta}$ (ABP parameters: $D=0.1$, $v=1.83$, $\mu=0.1$. Potential parameters: $k_x=6$, $k_y=20$).}\label{fig:ReactiveProbDensity_vs_Dr}
\end{figure}

In this section we study the ABP behavior for three different values of the P\'eclet number, $\text{Pe} := v \sqrt{3/4DD_{\vartheta}}$ obtained by keeping fixed the velocity $v=1.83$ (corresponding to $\text{Pe}=5$ in the main text) and changing the rotational diffusion parameter: $\text{Pe} = 3.5$ ($D_{\vartheta}=2$), $\text{Pe} = 7.1$ ($D_{\vartheta}=0.5$), and $\text{Pe} = 10$ ($D_{\vartheta}=0.25$).
Similarly to what happens when increasing the P\'eclet number by increasing the self-propulsion velocity $v$, the average TPT grows with the P\'eclet number and the TPT distribution becomes broader with more and more long-lasting trajectories (see Fig.~\ref{fig:Distribution_TPT_vs_Dr}).
However, an interesting new feature emerges. Namely, at a low rotational diffusivity the TPT distribution becomes bimodal (see Fig.~\ref{fig:Distribution_TPT_vs_Dr} for $\text{Pe} = 10$).
This bimodality shows that, an increase in the P\'eclet number obtained by decreasing the rotational diffusion rather than by increasing the velocity results in a more clear separation between short reactive paths traveling close to the minimum-energy path and long-lasting trajectories in which the particles ``surf'' along the energy walls before landing into the target.
The plots of the reactive probability density and the transition current further underline the differences of the ABP behavior when the same P\'eclet number is obtained through different sets of the parameters $v$ and $D_{\vartheta}$ (see Fig.~\ref{fig:ReactiveProbDensity_vs_Dr}).
For example, a comparison of Fig.~3c ($v=3.65$, $D_{\vartheta}=1$) and Fig.~\ref{fig:ReactiveProbDensity_vs_Dr}c ($v=1.83$, $D_{\vartheta}=0.25$), both corresponding to $\text{Pe}=10$, shows that when the velocity is higher the transition-path density is larger in the region behind the basins while when the rotational diffusivity is higher the the transition path density is larger in the regions below and above the basins.

\begin{figure}[h!]
\centering
\includegraphics[width=0.7\textwidth]{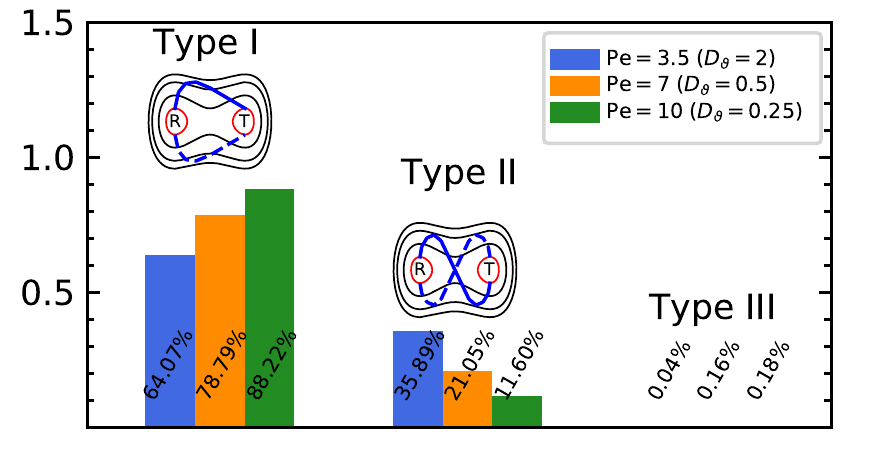}
\vspace{-10pt}
\caption{Frequency of different trajectory types for three values of the P{\'e}clet number obtained by changing $D_{\vartheta}$. A sketch of the trajectory type is given above the bars (ABP parameters: $D=0.1$, $v=1.83$, $\mu=0.1$. Potential parameters: $k_x=6$, $k_y=20$).}\label{fig:histogram_Dr}
\end{figure}

Similarly to what has been done in Section IV, different transition pathway are sorted in the three categories (see Fig.~\ref{fig:histogram_Dr}).
Again most of the trajectories are of type I with a relative frequency growing with the P{\'e}clet number.
Type II trajectories are the second most abundant type at each  P{\'e}clet number considered and very few other types of trajectories are observed.
Interestingly enough, a comparison with Fig.~\ref{fig:histogram} shows that at fixed P{\'e}clet number, $\text{Pe}=10$, type III trajectories are more frequent for the parameters set ($v=3.65$, $D_{\vartheta}=1$) than for ($v=1.83$, $D_{\vartheta}=0.25$).

\section{Behavior for strictly confining potential at $D=0$}

Up to now, in our study the self-propulsion velocity $v$ is always such that the barrier can be crossed even without the translational diffusion term.
However, similarly to what is done in Ref.~\cite{Woillez2019}, it is also interesting to consider another regime, namely the case in which the potential is a strictly confining potential in the limit $D\rightarrow0$, i.e. the barrier cannot be crossed in the absence of fluctuations.

\begin{figure}[h!]
\centering
\includegraphics[width=0.75\textwidth]{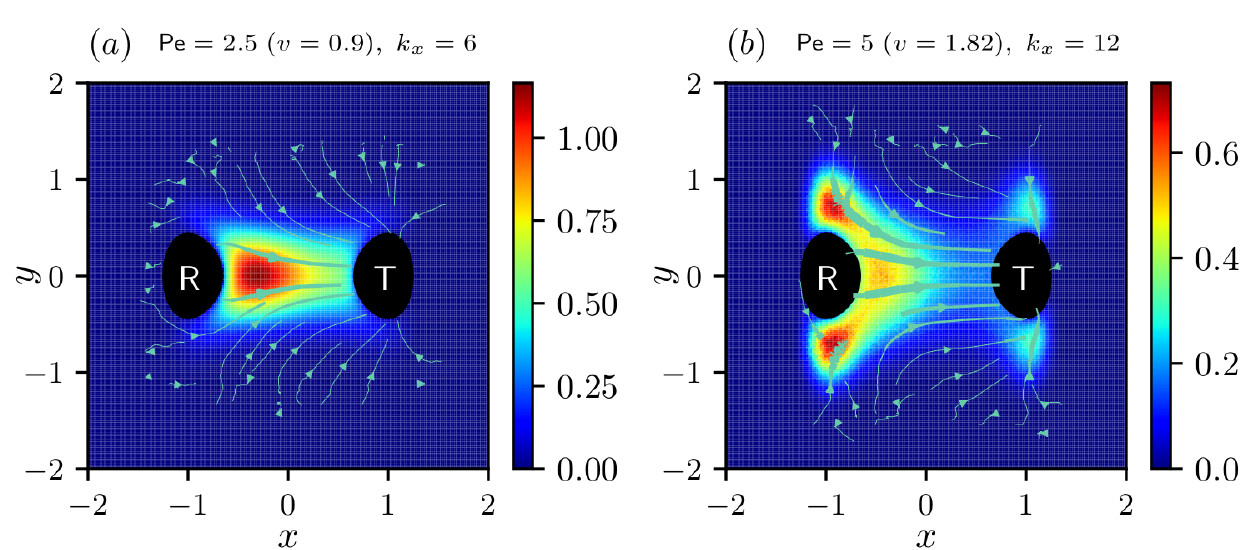}
\caption{$m(\vec{r})$ and $\vec{J}(\vec{r})$ in the reactive region for the parameters reported above the panels. (Fixed parameters: $k_y=20$, $D=0.1$, $D_{\vartheta}=1$, $\mu=0.1$).} \label{fig_mstrictlyconfpot}
\end{figure}

Assuming the potential landscape discussed in the main text and keeping fixed all others parameters, the maximum value of $v$ for which the potential is strictly confining at $D=0$ is $v=0.9$, corresponding to a P{\'e}clet number $\text{Pe} = 2.46$.
A plot of the reactive probability density and of the reactive current, shows that for such a low activity the behavior of the ABP is very similar to the behavior of a passive particle, i.e. its transition pathways travel close to the minimum free energy path without the long detours typical of more active particles (see Fig.~\ref{fig_mstrictlyconfpot}a).
Alternatively, one can consider a higher activity and make the potential strictly confining potentials at $D=0$ by increasing the stiffness parameter $k_x$.
For the case of $\text{Pe} = 5$ ($v=1.83$) considered in the main text this condition is obtained for $k_x = 12$.
In this case, one can notice that, while the reactive paths of the ABP tend to go through the saddle point of the energy landscape (in $x=0$, $y=0$), the way they exit the reactant basin is very different from the passive case (see Fig.~\ref{fig_mstrictlyconfpot}b).

\section{Crossing rates from reactant basin to target basin}

Table~\ref{tab:rates} reports the crossing rates from reactant basin to target basin in the various potential and parameter setups we have considered in our study.
Even if in principle it is possible to obtain information on rates in the TPS framework complemented with transition interface
sampling\cite{verp2003,vanerp2005}, at this stage we obtained the crossing rates directly from  brute force simulations.
From the Table it is possible to see how, for the parameter sets considered, the rates of active particles are always about two orders of magnitude higher than those of passive particles.
However, considering ABPs only, the dependence of rates on the parameters is less clear.
For example, the crossing rate grows with the P{\'e}clet number if the latter increases because of a change in the self-propulsion velocity, but it decreases if the increase in the P{\'e}clet number is due to a lower rotational diffusion coefficient.
We expect that the target-search process is optimized for some particular set of parameters and that this set depends on the particular energy landscape under considerations.
In particular, one may be interested in finding the optimal set of parameters in a complex energy landscape with several low-lying minima.

\begin{center}
\begin{table} \label{tab:rates}
\begin{tabular}{|c|c|c|c|c|c|c|}
\hline 
 & $\text{Pe}=0$ & $\text{Pe}=3.5$ & $\text{Pe}=5$ & $\text{Pe}=7$ & $\text{Pe}=10$ & $\text{Pe}=10$ \\  
 & {\footnotesize ($D_{\vartheta}\!\!=\!1$, $v\!\!=\!0$)} & {\footnotesize($D_{\vartheta}\!\!=\!2$, $v\!\!=\!1.83$)} & {\footnotesize($D_{\vartheta}\!\!=\!1$, $v\!\!=\!1.83$)} & {\footnotesize($D_{\vartheta}\!\!=\!0.5$, $v\!\!=\!1.83$)} & {\footnotesize($D_{\vartheta}\!\!=\!0.25$, $v\!\!=\!1.83$)} & {\footnotesize($D_{\vartheta}\!\!=\!1$, $v\!\!=\!3.65$)} \\ 
\hline 
{\footnotesize double well ($k_x\!\!=\!6$, $k_y\!\!=\!20$)} & $6.32 \cdot 10^{-4}$ & $5.61 \cdot 10^{-2}$ & $5.63 \cdot 10^{-2}$ & $4.46 \cdot 10^{-2}$ & $2.93 \cdot 10^{-2}$ & $5.62 \cdot 10^{-2}$ \\ 
\hline 
{\footnotesize double well ($k_x\!\!=\!8$, $k_y\!\!=\!20$)} & $1.15 \cdot 10^{-4}$ & $4.15 \cdot 10^{-2}$ & $4.6 \cdot 10^{-2}$ & $3.92 \cdot 10^{-2}$ & $2.7 \cdot 10^{-2}$ & $6.07 \cdot 10^{-2}$ \\ 
\hline 
{\footnotesize double well ($k_x\!\!=\!10$, $k_y\!\!=\!20$)} & $1.93 \cdot 10^{-5}$ & $2.59 \cdot 10^{-2}$ & $3.25 \cdot 10^{-2}$ & $3.1 \cdot 10^{-2}$ & $2.31 \cdot 10^{-2}$ & $6.25 \cdot 10^{-2}$ \\ 
\hline 
{\footnotesize M\"{u}ller-like potential} & $2.61 \cdot 10^{-4}$ & $3.99 \cdot 10^{-2}$ & $3.94 \cdot 10^{-2}$ & $3.05 \cdot 10^{-2}$ & $2.03 \cdot 10^{-2}$ & $4.85 \cdot 10^{-2}$ \\ 
\hline 
\end{tabular}  
\caption{Crossing rates from reactant basin to target basin. In all cases $D=0.1$ and $\mu=0.1$. Parameters of the M\"{u}ller-like potential are reported in Section~\ref{sec_differentU}. Rates are in units of the inverse computational time.}
\end{table}
\end{center}


\begin{thebibliography}{69}
\bibitem [{spe()}]{specification_passive_case}%
  \BibitemOpen
  \href@noop {} {}\bibinfo {note} {For a passive particle the mobility is
  related to the translational diffusion coefficient via the
  fluctuation-dissipation theorem $D = \mu k_B T$, with $T$ the temperature of
  the thermal bath in which the particle is immersed}\BibitemShut {NoStop}%
\bibitem [{len()}]{length_paths}%
  \BibitemOpen
  \href@noop {} {}\bibinfo {note} {Note that the two paths
  $\mathcal{W}^{\alpha}$ and $\mathcal{W}^{\beta}$ in general have different
  lengths and that consequently the index $i$ is unlikely to be equal to index
  $j$}\BibitemShut {NoStop}%
\bibitem [{sho()}]{shooting_point}%
  \BibitemOpen
  \href@noop {} {}\bibinfo {note} {Due to the stochastic nature of its
  dynamics, for an ABP the perturbation of the shooting point is not necessary.
  Hence, we set $w_{i}^{\beta} = w_{j}^{\alpha}$ with probability equal to
  one}\BibitemShut {NoStop}%
\bibitem [{lon()}]{long_TPT}%
  \BibitemOpen
  \href@noop {} {}\bibinfo {note} {Clearly, it takes a time of the order of the
  rotational diffusion time before the particle points again in a favorable
  direction, thus leading to long TPTs}\BibitemShut {NoStop}%
\end{thebibliography}
\end{document}